\documentclass[%
 reprint,
superscriptaddress,
showkeys,
nofootinbib,
 amsmath,amssymb,
 aps,
]{revtex4-2}
\usepackage{color, soul, framed}
\usepackage{multirow}
\usepackage{graphicx}% Include figure files
\usepackage{dcolumn}% Align table columns on decimal point
\usepackage{bm}% bold math
\usepackage{hyperref}% add hypertext capabilities
\usepackage[mathlines]{lineno}% Enable numbering of text and display math
\begin{document}

\preprint{APS/123-QED}

\title{High-throughput decoder of quasi-cyclic LDPC codes with \\ limited precision for continuous-variable quantum key\\ distribution systems}% Force line breaks with \\

\author{Chuang Zhou}
\affiliation{%
Science and Technology on Communication Security Laboratory, Institute of Southwestern Communication, Chengdu 610041, China}%
\author{Yang Li}
\email{yishuihanly@pku.edu.cn}
\affiliation{%
Science and Technology on Communication Security Laboratory, Institute of Southwestern Communication, Chengdu 610041, China}%
\author{Li Ma}
\author{Jie Yang}
\author{Wei Huang}
\author{Heng Wang}
\author{Yujie Luo}
\affiliation{%
Science and Technology on Communication Security Laboratory, Institute of Southwestern Communication, Chengdu 610041, China}%
\author{Francis C. M. Lau}
\affiliation{%
Department of Electronic and Information Engineering, The Hong Kong Polytechnic University, Hong Kong}%
\author{Yong Li}
\email{yongli@cqu.edu.cn}
\affiliation{%
College of Computer Science, Chongqing University, Chongqing 400044, China}%
\author{Bingjie Xu}
 \email{xbjpku@pku.edu.cn}%
 \affiliation{%
Science and Technology on Communication Security Laboratory, Institute of Southwestern Communication, Chengdu 610041, China}%
\date{\today}% It is always \today, today,

\begin{abstract}
Abstract— More than Mbps secret key rate was demonstrated for continuous-variable quantum key distribution (CV-QKD) systems, but real-time postprocessing is not allowed, which is restricted by the throughput of the error correction decoding in postprocessing. In this paper, a high-throughput FPGA-based quasi-cyclic LDPC decoder is proposed and implemented to support Mbps real-time secret key rate generation for CV-QKD for the first time. A residual bit error correction algorithm is used to solve the problem of high frame errors rate (FER) caused by the limited precision of the decoder. Specifically, real-time high-speed decoding for CV-QKD systems with typical code rates 0.2 and 0.1 is implemented on a commercial FPGA, and two throughputs of 360.92Mbps and 194.65Mbps are achieved, respectively, which can support 17.97 Mbps and 2.48 Mbps real-time generation of secret key rates under typical transmission distances of 25km and 50km, correspondingly. The proposed method paves the way for high-rate real-time CV-QKD deployment in secure metropolitan area network.

\end{abstract}

%Use showkeys class option if keyword
\maketitle

%\tableofcontents

\section{\label{sec:level1}INTRODUCTION}
Based on the principle of quantum mechanism, quantum key distribution (QKD) can share unconditional security keys between two communication parties under insecure channels \cite{r1,r2}. According to the modulation methods adopted in  quantum state exchanges, QKD protocols are usually divided into two categories: discrete-variable quantum key distribution (DV-QKD) and continuous-variable quantum key distribution (CV-QKD). CV-QKD has attracted much attention and has made great breakthroughs in recent years \cite{r3,r4,r5} by utilizing the existing optical communication infrastructure.

The CV-QKD protocol is comprised of two main procedures: quantum state transmission over a quantum channel and classical postprocessing over an authenticated classical public channel. After quantum state transmission, the communication parties share unidentical correlated data. Then, postprocessing is implemented to transform the correlated data to identical secure keys. 

In recent years, the secure key rate (SKR) of CV-QKD has been greatly improved by robustly increasing the repetition rate of modulation and detection of coherent states \cite{r6,r7,r8,r9,r10}. As two main CV-QKD schemes based on Gaussian modulation coherent states \cite{r11,r12} and discrete modulation coherent states \cite{r13,r14}, the repetition rates of their transceivers have been increased to above 100MHz, which can support the distribution of tens of Mbps SKR. Especially, in the latest work, a four-state discrete-modulated CV-QKD with a 5GBaud repetition rate has been experimentally demonstrated for further increasing the asymptotic SKR to sub-Gbps level \cite{r15}. However, most of reported SKRs are only obtained by simple theoretical calculation or off-line postprocessing, which is not real-time CV-QKD. The major difficulty is that the throughput of the postprocessing cannot support the real-time SKR generation in high repetition rate CV-QKD. Therefore, the development of high throughput postprocessing is of great importance for the practical application of high-rate and real-time CV-QKD.

The common postprocessing mainly includes four steps: data sifting, parameter estimation, information reconciliation and privacy amplification. For a practical CV-QKD system, information reconciliation has an important influence on the secret key rate.

For information reconciliation of CV-QKD, a specific error-correcting code is adapted to correct errors in sifted keys to obtain identically corrected keys. Since a CV-QKD system usually works in a Gaussian channel with very low signal-to-noise ratios (SNRs), typically below 0dB, commercial error-correcting codes are not suitable for CV-QKD. Several error-correcting codes are studied for CV-QKD, such as LDPC codes \cite{r16,r17}, polar codes \cite{r18,r19}, and raptor codes \cite{r20,r21,r22}. Among them, a special class of LDPC codes, i.e., multi-edge-type low density parity-check (MET-LDPC) codes, is chosen as the most promising error-correcting code for CV-QKD, because its performance is very close to Shannon capacity limit and the decoding algorithm can be accelerated by using devices with parallel processing capabilities. MET-LDPC codes with low code rate and long code length are designed for CV-QKD \cite{r23}, and high-speed implementation of the decoding algorithm for MET-LDPC codes based on graphic processing unit (GPU) is shown in \cite{r17} with a throughput of 9.17 Mbps. Multiple codewords decoding in parallel is proposed to improve the throughput to 30.39 Mbps \cite{r24}, and layered decoder for Quasi-cyclic MET-LDPC (QC MET-LDPC) are proposed to further improve decoding throughput to 64.11 Mbps \cite{r25}.

Although high decoding speed of MET-LDPC codes can be achieved by using GPU, the power consumption, volume, and cost of GPU limit its application in practice. FPGA is a kind of large-scale programmable integrated device and has a series of advantages, such as parallelism, high processing speed, and low power consumption. Many FPGA-based LDPC decoders have been developed over past few decades \cite{r26,r27,r28}. Recently, FPGA-based MET-LDPC decoders for CV-QKD are also studied \cite{r29,r30} and the fixed-point number of width 19 is adopted in designing the decoder. However, according to our knowledge, a successful high-speed LDPC decoder based on FPGA for a CV-QKD system under a typical transmission distance, such as 50 km, has not been realized in the literature.

The difficulty of a high-speed FPGA-based decoder for CV-QKD is caused by two reasons. On the one side, the on-chip hardware resource of FPGA is limited, and a finite width (precision) of the fixed-point number is required for a high-speed decoder. On the other hand, the SNR for a CV-QKD under a typical transmission distance is much less than 1, so the decoding performance is strongly associated with the decoding accuracy. However, the limited precision decreases the decoding accuracy, leading to the existence of residual error-bits after decoding and high frame errors rate (FER), which decreases the secret key rate drastically.

In this paper, a FPGA-based decoder for CV-QKD with limited precision is studied. The characteristics of the residual error-bits are studied, and it is found that a self-defined reliability value can be used to distinguish the error bits from the correct bits \cite{r31}. So, a residual bit error correction algorithm based on reliability values is used to improve the decoding performance. The improved decoding algorithm is implemented on FPGA, and  throughputs of 360.92Mbps and 194.65Mbps are achieved for code rates 0.2 and 0.1 respectively, which can support 17.97 Mbps and 2.48 Mbps real-time generation of secret key rates under transmission distance of 25km and 50km, correspondingly. The proposed method can support Mbps real-time secret key rate generation for CV-QKD, which paves the way for a large-scale deployment of high-rate real time CV-QKD systems in secure metropolitan area network.

The rest of this paper is organized as follows. Section ~\ref{sec:level2} gives a brief introduction into QC-LDPC codes and the iterative-decoding algorithm. The proposed architecture of the decoder and its residue error-bits erase module are introduced in Section ~\ref{sec:level3}. Implementation results are discussed in Section ~\ref{sec:level4} and we analyze the corresponding secret key rate of the decoder for a CV-QKD system in Section ~\ref{sec:level5}. Finally, Section ~\ref{sec:level6} draws a conclusion.

\section{\label{sec:level2} ERROR CORRECTING SCHEME}
\subsection{\label{sub:level1}QC-LDPC}
\begin{figure}[h]
\centerline{
\includegraphics[width = 8cm]{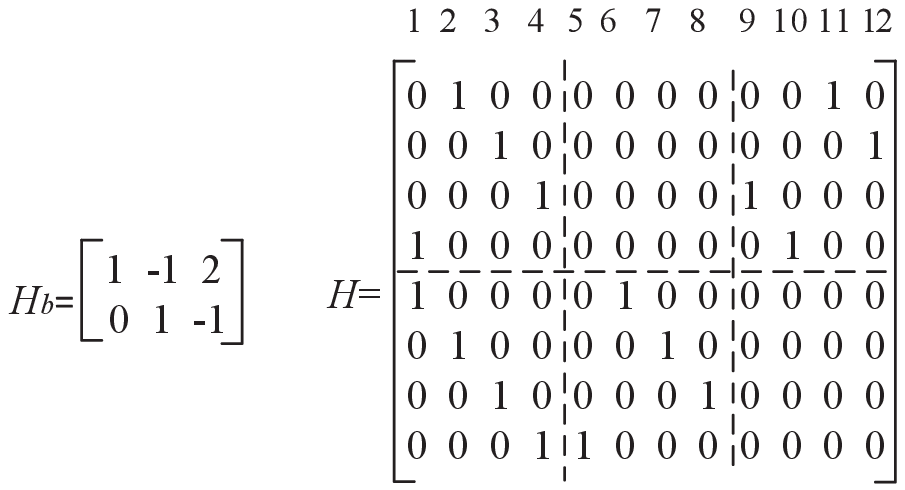}}% Here is how to import EPS art
\caption{\label{fig1}Generating \emph{\textbf{H}} from \emph{\textbf{H}}$_{b}$. The size of \emph{\textbf{H}}$_{b}$ is $2\times3$, the size of CPM is $4\times4$, and the size of H is $8\times12$. The number 1, 2, … ,12 correspond to the column-index of \emph{\textbf{H}}.}
\end{figure}

QC-LDPC codes \cite{r32} is a class of structured LDPC codes. When implementing the decoding algorithm in FPGA, the QC structure can reduce its  complexity of implementation. The parity-check matrice \emph{\textbf{H}} of a QC-LDPC code can be generated by using a cyclic permutation matrix (CPM) to replace the elements in base matrix \emph{\textbf{H}}$_{b}$. If an element in \emph{\textbf{H}}$_{b}$ is ‘$-1$’, the CPM used to replace this element is a $Z\times{Z}$ zero matrix. Matrix used to replace the non-negative integer $\alpha$  in
\emph{\textbf{H}}$_{b}$ was obtained by cyclically shifting the $Z\times{Z}$ identity matrix to the right by $\alpha$ bit. An example of generating \emph{\textbf{H}} from \emph{\textbf{H}}$_{b}$ with $Z=4$ is shown in Fig.~\ref{fig1}.

\subsection{\label{sub:level1}Decoding algorithm}
In the proposed architecture, the layered BP decoding algorithm for QC LDPC codes\cite{r20} is employed, which  achieves almost the same performance as the flooding BP algorithm with less decoding iterations. We denote  \emph{\textbf{u}} as a codeword over the BIAWGN channel and \emph{\textbf{R}} as the received vector. Layered BP decoding algorithm is described as follows.\\
(a)Initialize the LLR of variable nodes.
\begin{equation}\label{eq1}
\begin{aligned}
L_{q_{n} }^{(0,0)} =\frac{2R_{n} }{\sigma ^{2} },
\end{aligned}
\end{equation}

where $n$ is the variable node index, and $\sigma ^{2}$ denotes the variance of noise in the BIAWGN channel.\\
(b)Update the message transmitted from variable nodes to check nodes.
\begin{equation}\label{eq2}
\begin{aligned}
L_{q_{nm} }^{(t,l)}=L_{q_{n} }^{(t,l-1)}-L_{r_{mn} }^{(t-1,l)},
\end{aligned}
\end{equation}

where $ m$ is the check node index, $t$ is the number of iterations and $l$ is the number of layers, when $t=1$, $L_{r_{mn} }^{(0,l)}=0$.\\
(c)Update the message transmitted from check nodes to variable nodes.
\begin{equation}\label{eq3}
\begin{aligned}
L_{r_{mn} }^{(t,l)}=(1-2s_{m})\times \prod_{n^{'}\in N(m)\setminus n }^{} sgn(L_{q_{n^{'} } }^{(t-1,l)})\\\times \Phi^{-1}
(\sum_{n^{'}\in N(m)\setminus n}^{}\Phi(|L_{q_{n^{'} } }^{(t-1,l)}|)),
\end{aligned}
\end{equation}

where $\Phi(x)=\Phi^{-1}(x)=-ln(tanh(x/2))$ , $sgn(x)$ is a sign function, and $N(m)$ is the set of variable nodes connected to the check node $m$. Moreover, \emph{\textbf{s}}  is the syndrome generated by\emph{\textbf{ s}}=\emph{\textbf{uH}}$^{T}$.\\
(d)Update the total message of Variable nodes
\begin{equation}\label{eq4}
\begin{aligned}
L_{q_{n} }^{(t,l)}=L_{q_{nm} }^{(t,l)}+L_{r_{mn} }^{(t-1,l)}.
\end{aligned}
\end{equation}
Repeat steps (b), (c), (d) until all sub-matrices are updated and reach the maximum number of iterations. When all the layers finish updating message, $t = t + 1$. \\
(e)Decide the codeword $\hat{ \emph{\textbf{u}}}$ according to the total message of variable nodes.
\begin{equation}\label{eq5}
\begin{aligned}
{\hat{u}}_{n} =\left\{\begin{matrix}
 1, L_{q_{n} }^{(t,l)} <0 \\
0, L_{q_{n} }^{(t,l)} \ge 0
\end{matrix}\right.
.
\end{aligned}
\end{equation}

\section{\label{sec:level3} PROPOSED ARCHITECTURE}
\begin{figure*}[t]
\includegraphics[width = 15cm]{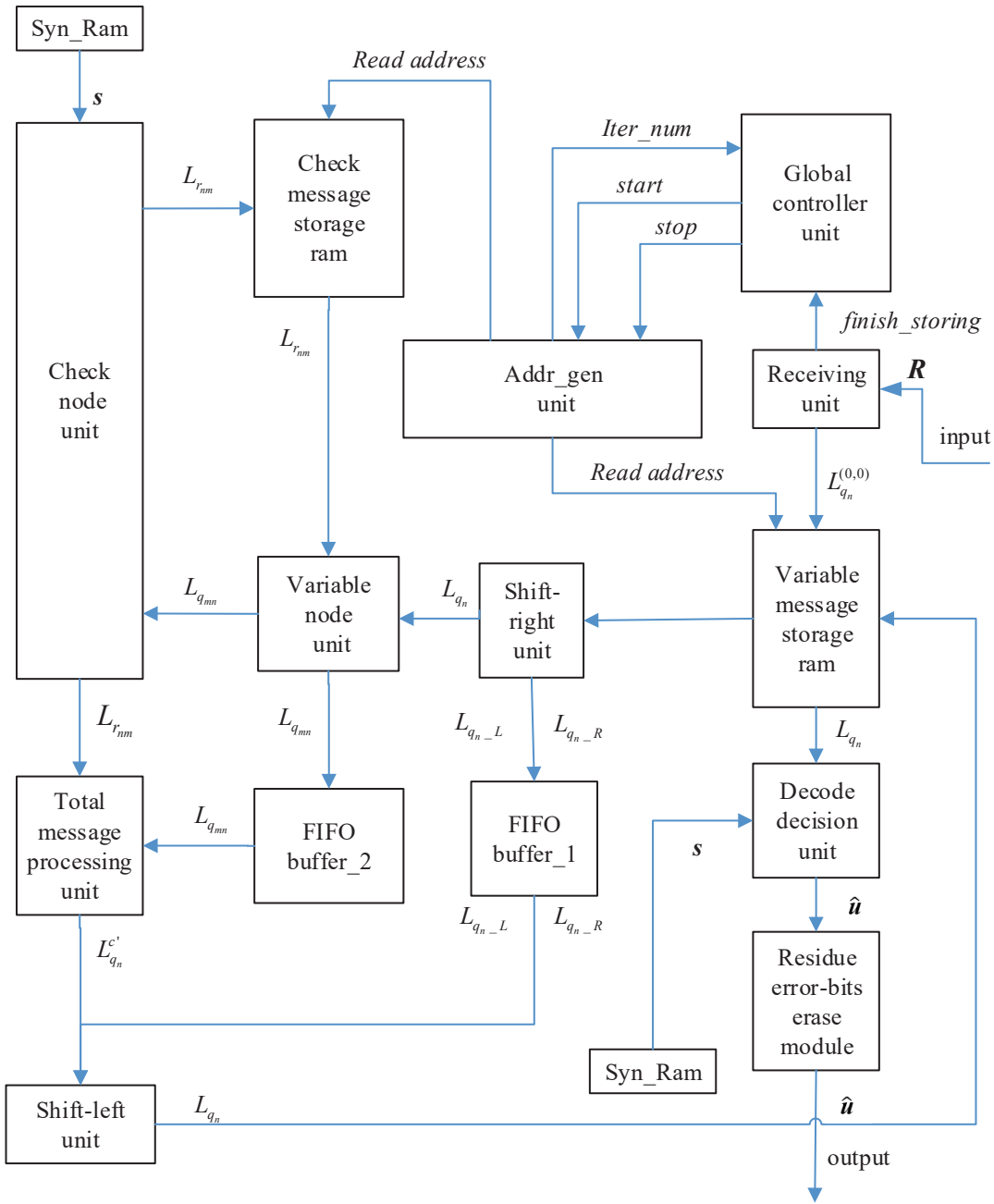}% Here is how to import EPS art
\caption{\label{fig2}The proposed architecture of FPGA-based LDPC decoder. \emph{\textbf{R}} is the data that received from the channel. The global controller unit deals with the whole procedure of decoding and $\hat{ \emph{\textbf{u}}}$. is the decoding result.}
\end{figure*}

As shown in Fig.~\ref{fig2}, the proposed architecture of the decoder contains a global controller unit (GCU), a receiving unit, a Addr$\_$gen unit, a decode decision unit, two FIFO buffers, four processing units, three types of RAMs and a residue error-bits erase module.  

The global controller unit deals with the whole procedure of decoding. The receiving unit receives the message that needs to be decoded and stores it into the variable message storage RAM with right order.

The variable message storage RAM is used to store the value of $L_{q_{n} }$ in Eqs.(~\ref{eq1}) and (~\ref{eq4}). The check node message storage RAM is used to store the value of $L_{r_{mn} }$ in Eq.(~\ref{eq3}). The Syn$\_$RAM is used to store \emph{\textbf{s}} in  Eq.(~\ref{eq3}). Besides, in the decode decision unit, Eq.(~\ref{eq5}) is executed and we get the vector $\hat{ \emph{\textbf{u}}}$ and its syndrome $\hat{ \emph{\textbf{s}}}$ by computing $\hat{ \emph{\textbf{u}}}$ =  $\hat{ \emph{\textbf{s}}}$\emph{\textbf{H}}$^{T}$. Then, we read \emph{\textbf{s}} from Syn$\_$RAM. If \emph{\textbf{s}} is equal to $\hat{ \emph{\textbf{s}}}$, the residue error-bits erase module just outputs $\hat{ \emph{\textbf{u}}}$. Otherwise, the residue error-bits erase module starts erasing the residual error-bits in $\hat{ \emph{\textbf{u}}}$ and then outputs $\hat{ \emph{\textbf{u}}}$.

\subsection{\label{sub:level1}Datapath}
The receiving unit output $finish$\_$storing$ represents it finishes storing data into the variable message storage RAM. Then, the global controller unit uses signal $start$ to control the addr$\_$gen unit to generate read address of the variable message storage RAM and the iterative decoding starts.

Data $L_{q_{n}}$ is read from the variable message storage RAM. By using the Shift-right unit, we will get the data that we need in the current clock period. Then, the VNU is used to execute Eq.(~\ref{eq2}). The output of VNU is used as the input of the CNU and is stored in the FIFO buffer$\_$2 at the same time. The function of Eq.(~\ref{eq3}) is realized at the CNU and the output of this unit is used as the input of the total message processing unit and is written back to the check message storage RAM. The total message processing unit executes  Eq.(~\ref{eq4}) and its output is processed by the shift$\_$left unit to get the right order of $L_{q_{n}}$. Next, $L_{q_{n}}$ will be written back to the variable message storage RAM. After finishing writing, an iteration is completed, and the value of $Iter$\_$num$ adds one. When the value of $Iter$\_$num$ is equal to the maximum number of iterations, the decoding stops. 

\subsection{\label{sub:level2}Fixed-point number}
When implementing the decoder based on FPGA, we adopt fixed-point number to reduce the implementation complexity. The width of the fixed-point number \emph{w} = 1 + \emph{I} + \emph{F}, one bit for the sign, \emph{I} bit for the integer part and \emph{F} bit for the fraction part. In the design of the decoder, we can reduce the consumption of hardware resources by reducing the width of the fixed-point number.
\subsection{\label{sub:level3}Partial parallel and pipeline}
Parallelism is useful to improve the decoding throughput. However, restricted by the resource of the FPGA, the parallelism is limited. An MET-LDPC check matrix with a proper quasi-cycle is chosen to make full use of the parallelism.

Pipeline structure is often adopted with the parallelism to improve the throughput. We can minimize the number of clock cycles required in the overall decoding process by using this structure. Because there is delay between reading data from the variable message storage RAM and writting them back, a suitable parity check matrix is needed to avoid read/write conflicts for pipeline structure.

\subsection{\label{sub:level4}Residue error-bits erase module}
As mentioned above, we can reduce the consumption of hardware resources by reducing the width of the fixed-point number. However, reducing the width of the fixed-point number also decreases its accuracy, which leads to the decrease of the decoding performance.

\begin{figure}[htb]
\centerline{
\includegraphics[width = 9cm]{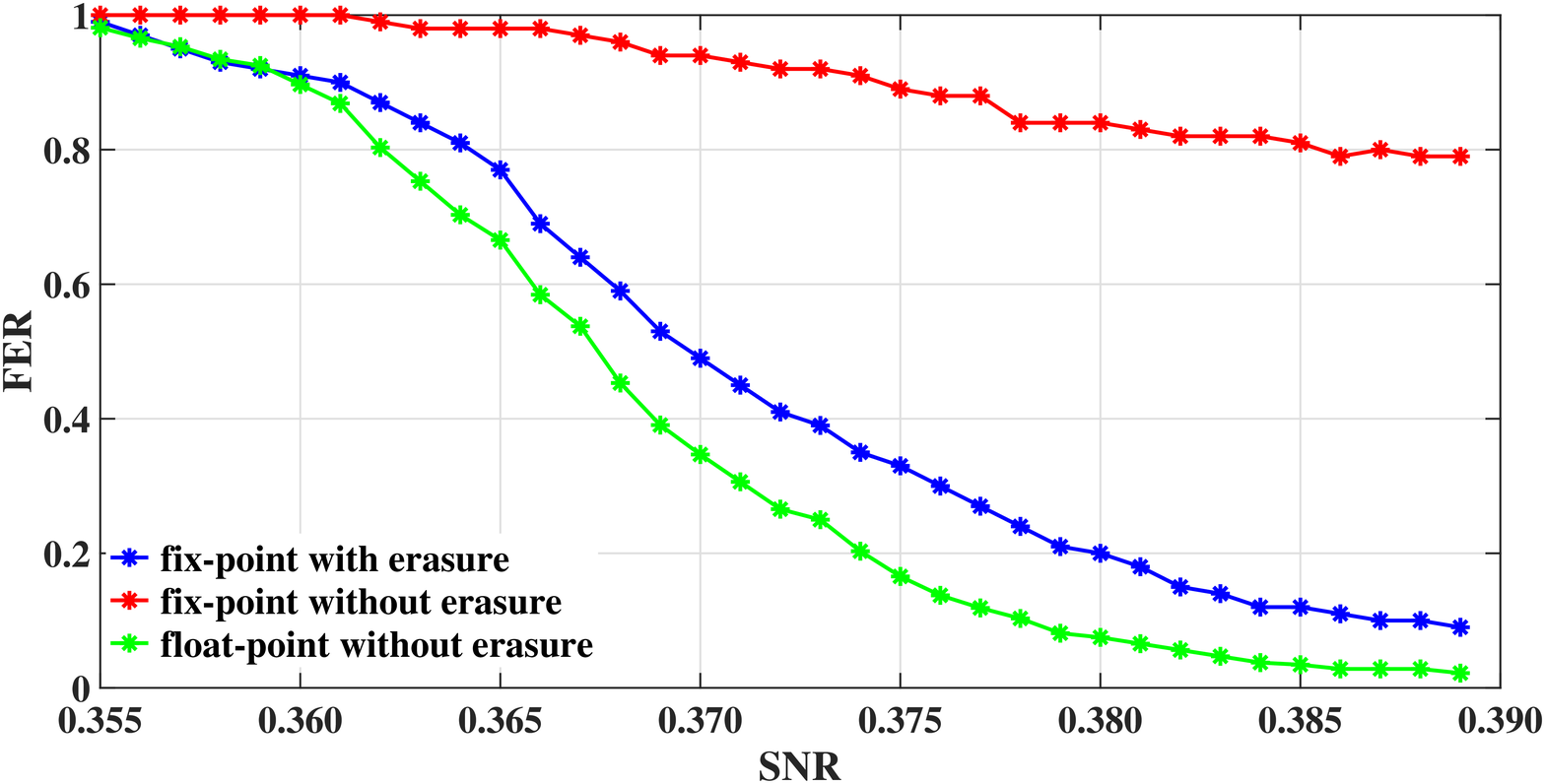}}% Here is how to import EPS art
\caption{\label{fig3} The FER vs SNR curves of the decoder for a rate-0.2 QC-LDPC code. The code length is 80000. The range of SNR in this simulation is [0.355, 0.389] and the interval is 0.001. For each SNR, the decoding result of 100 codewords are collected under $t_{max}$ = 13 , and the red/blue curve shows the decoding result of this decoder without/with residue error-bits erase module. The green curve shows the decoding result of this code when float-point number is adopted in the iterative decoding.}
\end{figure}

A rate-0.2 QC-LDPC code is taken for example. The width of the fixed-point number is set as \emph{w} = 8 (one symbol bit, 4 integer bits and 3 fraction bits) and the maximum number of decoding iterations  $t_{max}$ = 13  is adopted to design the decoder. As shown in Fig.~\ref{fig3}. The red curve shows the performance of this decoder with 100 frames are decoded at each SNR. We can see that the FER is extremely high, even if in the region of high SNR. The green curve shows the decoding performance of this code when float-point number is adopted in the iterative decoding, and we can see the FER decreases obviously. It should be mentioned that the high FER is not restricted by decoding iterations. So, the FER can’t be reduced just by increasing the iterations. 

\begin{figure}[htb]
\centerline{
\includegraphics[width = 9cm]{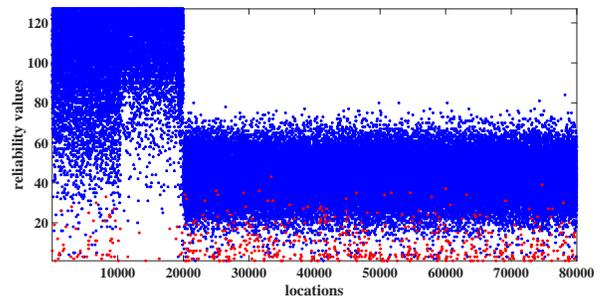}}% Here is how to import EPS art
\caption{\label{fig4} The reliability values of the LDPC codes after finishing decoding. The code rate is 0.2 and the code length is 80000. Four failed codewords are collected at SNR = 0.37. The erroneous symbols are marked with red asterisk and the correct ones with blue asterisk.}
\end{figure}

In order to improve the decoding performance of this design, we firstly analyze the decoding results. Reliability value is defined as the absolute value of $L_{q_{n}}$ in Eq.(~\ref{eq5}). When SNR = 0.37, four failed codewords are analyzed and the corresponding reliability values of these decoded symbols are shown in Fig.~\ref{fig4}. The erroneous symbols are marked with red asterisk and the correct ones with blue asterisk. It can be found that the reliability values of erroneous symbols are relatively smaller than those of the correct ones. To further reveal the feature of these decoded symbols, the histogram about the correct symbols and erroneous symbols are drawn in Fig.~\ref{fig5}(a) and Fig.~\ref{fig5}(b).

Hence, for a failed codeword, we can set a threshold $\Delta$. If the reliability value of a symbol is smaller than $\Delta$, it is added to the set of suspicious symbols, which is an erroneous symbol with high probability.

For a rate-0.2 QC-LDPC code, we set the threshold $\Delta$ to be 40 (the value of $L_{q_{n}}$ is in the range [-127, 127] when $w = 8$). For each SNR in Fig.~\ref{fig3}, 100 codewords are decoded. Parameter $N$\_$err$ is defined as the number of failed codewords whose maximum reliability value of erroneous symbols is smaller than $\Delta$, which is shown in Table~\ref{tab:tab1}. It can be found that most failed codewords are included, especially for relatively large SNRs. 

\begin{figure}[htb]
    {\label{fig5a}{\includegraphics[width = 9cm]{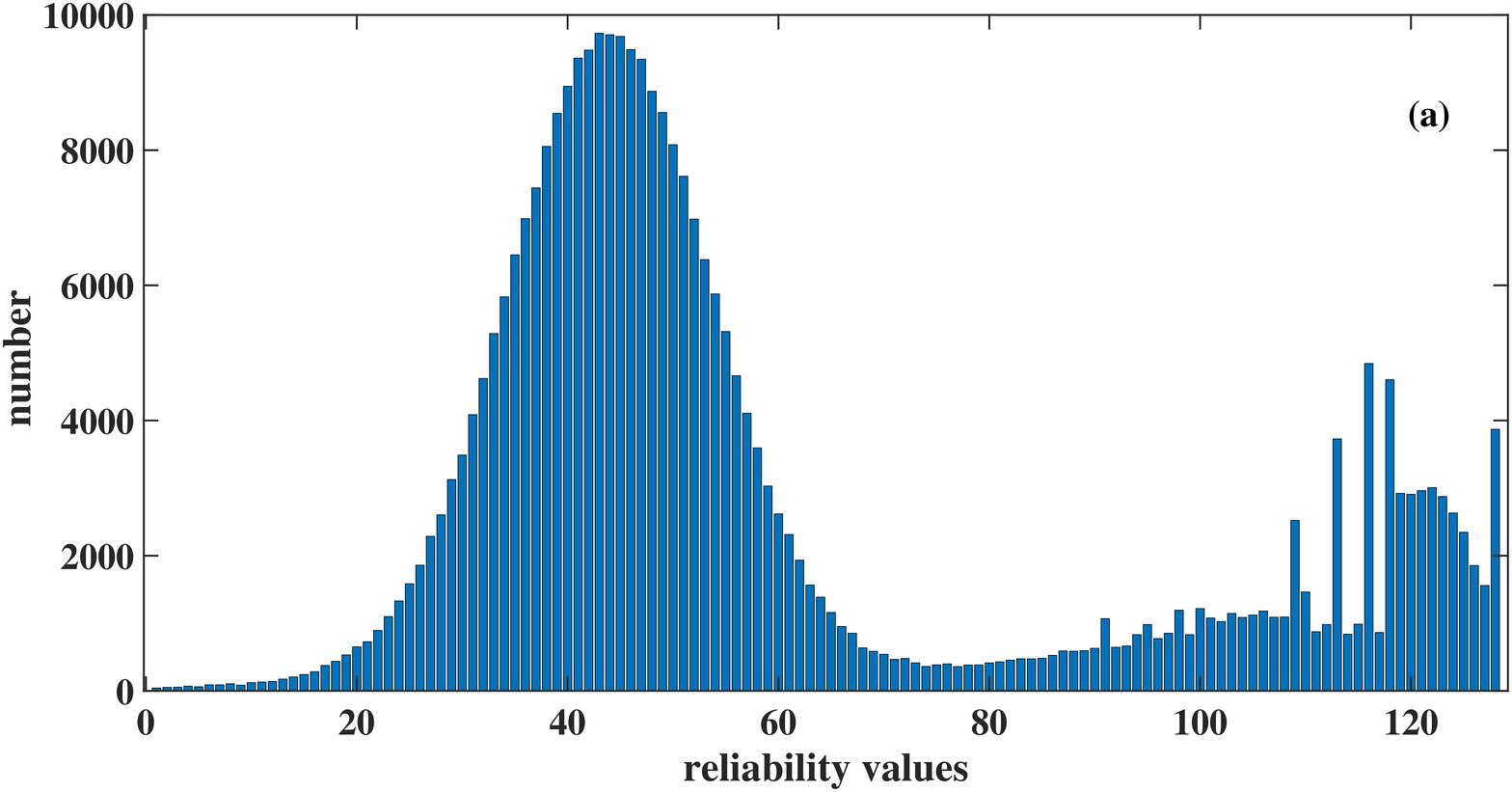}}}
    {\label{fig5b}{\includegraphics[width = 9cm]{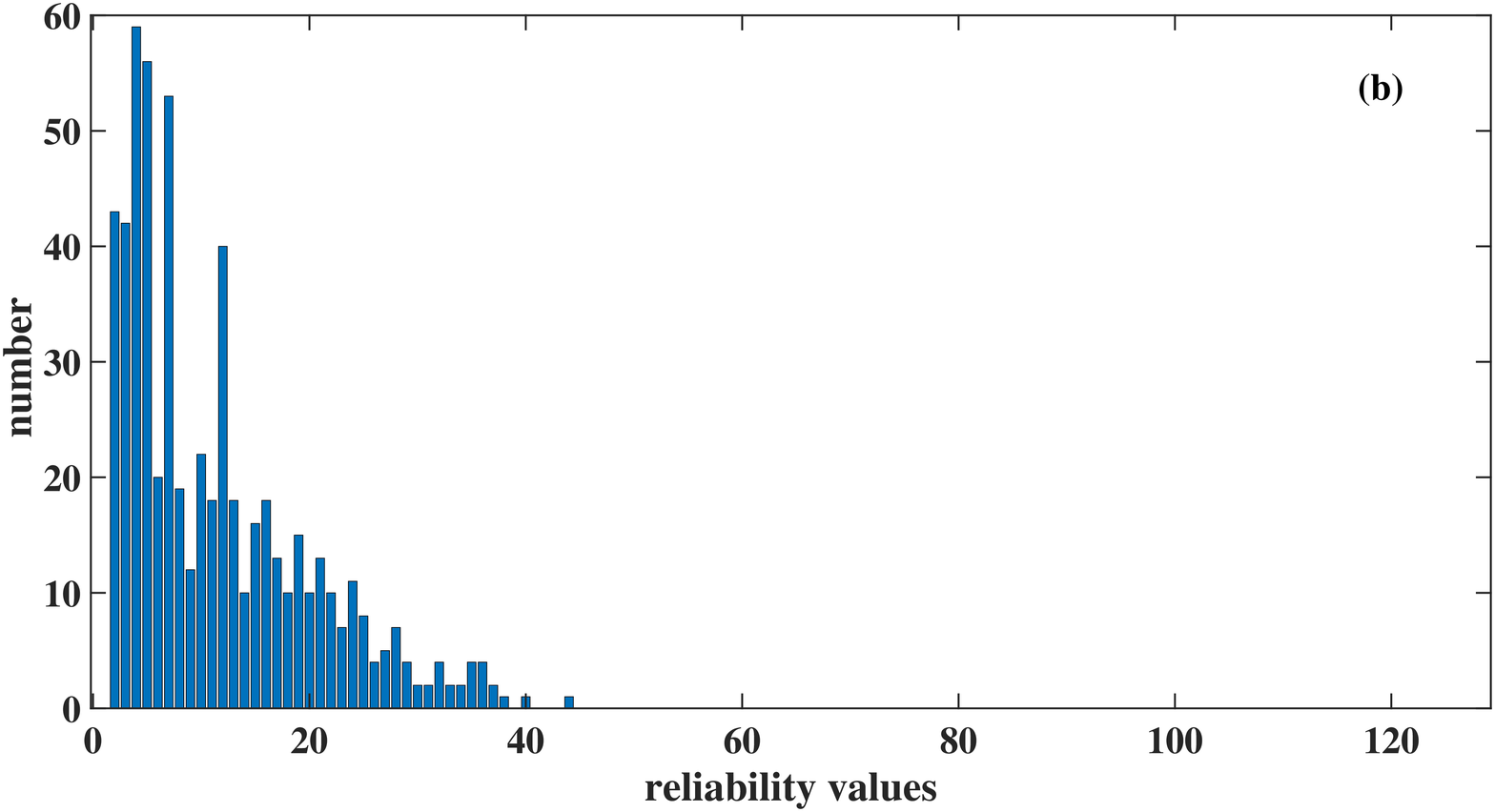}}}
    \caption{\label{fig5}Histogram of decoded symbols. (a): correct symbols; (b): wrong symbols. Four failed codewords are collected at SNR = 0.37.}{\label{fig5}}
\end{figure}

\begin{table*}[htb]
\caption{\label{tab:tab1}The number of failed codewords and the value of $N$\_$err$. For each SNR, the decoding result of 100 codewords are collected with $t_{max}$ = 13 and $\Delta$ = 40.}
\begin{ruledtabular}
\begin{tabular}{ccccccccc}
SNR & $0.368$ & $0.371$ & $0.374$ & $0.377$&$0.380$ &$0.383$& $0.386$& $0.389$\\ \hline
Number of failed codewords&$97$&$93$&$91$&$88$&$84$ &$82$&$79$&$79$ \\
$N$\_$err$&$44$&$51$&$59$&$63$&$64$ &$79$&$68$&$70$ \\

\end{tabular}
\end{ruledtabular}
\end{table*}

The distribution of reliability values is similar to the description in \cite{r31}, which is caused by the trapping set. In \cite{r31}, the trapping set influences the error floor of LDPC codes at high SNR. For our situation, the high FER occurs not only in the error floor region, but also in the waterfall region. It is not caused by the trapping set, but by the precision of the decoder. However, we can also try to use the method in \cite{r31} to erase the residual error-bits and we  implemente this method in the residue error-bits erase module.

When the LDPC decoder with residue error-bits erase module fails decoding, this module starts running to erase the residual error-bits. The blue curve in Fig.~\ref{fig3} shows the final decoding performance when $\Delta$ is set to be 40  in the residue error-bits erase module. We can observe that the decoding performance is improved greatly.

If fact, different $\Delta$ can be chosen for different SNRs. By a large amount of simulation, the performance improvement due to modification of $\Delta$ is not obvious. Therefore, in this paper, fixed $\Delta$ for different SNRs is adopted for simplicity.

\begin{figure}[htb]
\centerline{
\includegraphics[width = 9cm]{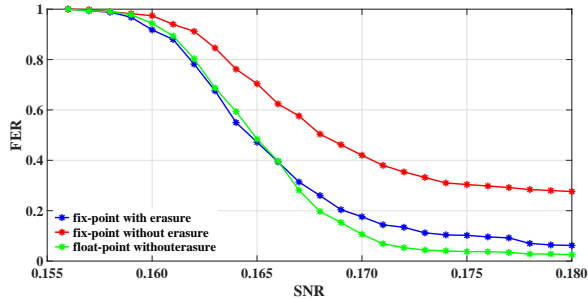}}% Here is how to import EPS art
\caption{\label{fig6} The FER vs SNR curves of the decoder for a rate-0.1 QC-LDPC code. The code length is 96000. The range of SNR in this simulation is [0.155, 0.180] and the interval is 0.001. For each SNR, the decoding result of 100 codewords are collected under  $t_{max}$ = 25, and the red/blue curve shows the decoding result of this decoder without/with residue error-bits erase module. The green curve shows the decoding result of this code when float-point number is adopted in the iterative decoding.}
\end{figure}

In a similar way, for a rate-0.1 QC-LDPC code, we set $w$ = 10 (one symbol bit, 4 integer bits and 5 fraction bits) and $\Delta$ = 180 (the value of $L_{q_{n}}$ is in the range [-511, 511] when $w$ = 10). The decoding performance (the FER vs SNR curve) of this decoder with/without residue error-bits erase module are studied. As shown in Fig.~\ref{fig6}, it can be seen that the residue error-bits erase module can also improve the decoding performance extremely for the considered codes.

\section{\label{sec:level4}IMPLEMTENTATION RESULTS}
In this section, the implementation results of the FPGA-based LDPC decoder are shown. A Xilinx VC709 evaluation board with a Virtex-7 XC7VX690T FPGA is chosen to implement the decoder. 
\subsection{\label{sub:level1}Throughput}
The throughput of the decoder is a bottleneck for the real-time performance of CV-QKD system. As a result, higher throughput is the aim of designing a FPGA-based decoder. The decoding throughput can be estimated by 
\begin{equation}\label{eq6}
\begin{aligned}
T=\frac{f\cdot N}{K\cdot t_{max} }, 
\end{aligned}
\end{equation}
where $f$ is the frequency of clock on FPGA, $N$ is code length, $K$ is the clock cycles that are used to finish one iteration decoding and $t_{max}$  is the maximum number of decoding iterations.
Because  we adopt pipeline structure to design the decoder, parameter $K$ is roughly equivalent to the clock cycles that are used to read data from the variable message storage RAM. We use $N_{total}$  to represent the number of nonzero elements in \emph{\textbf{H}} and $N_{var}$  to represent the average row-weight in \emph{\textbf{H}}. Then, the decoding throughput can be described as \cite{r30}
\begin{equation}\label{eq7}
\begin{aligned}
T\approx \frac{f\cdot N}{(N_{total}/p )t_{max} } \approx \frac{f\cdot N\cdot p}{N\cdot (1-R)\cdot N_{avr}\cdot t_{max}  } \\
\approx \frac{f\cdot p}{(1-R)\cdot N_{avr}\cdot t_{max}},  
\end{aligned}
\end{equation}
where $p$ is the parallelism, $R$ is the code rate of the LDPC codes. From Eq.(~\ref{eq7}), we can see that when given an LDPC code, the throughput can be improved by using higher clock frequency, enlarging parallelism parameter $p$ and reducing the maximum number of decoding iterations.

To decrease the FER of this proposed decoder, we add a module to erase the error bits after the iterative decoding. $D_e$ is used to represent the clock cycle delay of the residue error-bits erase module. Eq.(~\ref{eq7}) can be modified as
\begin{equation}\label{eq8}
\begin{aligned}
T\approx \frac{f\cdot p}{(1-R)\cdot N_{avr}\cdot t_{max}+D_e}.  
\end{aligned}
\end{equation}
\subsection{\label{sub:level2}Decoding performance}
In the following, we investigate the performance of two QC-LDPC codes with the code length of 96000 and 80000 and code rate of 0.1 and 0.2 by using the decoding algorithm proposed in section~\ref{sec:level2}. Table~\ref{tab:tab2} presents the parameters of the FPGA–based LDPC decoder for two different QC-LDPC codes and compares them with previous work.

\begin{table*}[htb]
\caption{\label{tab:tab2}Performance of the proposed LDPC decoder and its comparison with previous work.}
\begin{ruledtabular}
\begin{tabular}{ccccccccc}
Decoder  & \multicolumn{2}{c}{Yang. et al.\cite{r30}}&\multicolumn{2}{c}{This work} \\ \hline
FPGA device &\multicolumn{2}{c}{Virtex-7 XC7VX690} &\multicolumn{2}{c}{Virtex-7 XC7VX690} \\ 
Algorithm &\multicolumn{2}{c}{ Layered BP decoding} &\multicolumn{2}{c}{ Layered BP decoding} \\ 
Code rate &0.43	 &0.115 &0.1	 &0.2\\
Parallel parameter $p$	 &64	 &64	 &100	 &100\\
Width of fixed-point number(bits) &19	 &19	 &10	 &8\\
BRAMs of One decoder (kb)	&31680(59.9$\%$)	&30636 (57.9$\%$)	&22863 
(43.21$\%$)	&18144 (34.29$\%$)\\
Residue error-bits erase module	&No	&No	&Yes	&Yes\\
Decoder numbers	&1	&1	&2	&2\\
Clock frequency (MHz)	&100	&100	&100	&100\\
Throughput (Mbps)	&108.64	&70.32	&194.65	&360.92\\

\end{tabular}
\end{ruledtabular}
\end{table*}

\begin{table*}[htb]
\caption{\label{tab:tab3}The optimization results of secret key rate. Parameter $G_K$ represents the improvements of $K_{opt}$ by using the residue error-bits erase module.}
\begin{ruledtabular}
\begin{tabular}{ccccccc}
$R$  & $d$(km)&Residue error-bits erase module &$V_A$ &SNR&$K_{opt}$(bits/{pulse})&$G_K$\\ \hline
\multirow{2}{*}{0.2}	&\multirow{2}{*}{25}	&Yes	&6.3692	&0.3707	&0.0498	&\multirow{2}{*}{425.64$\%$}\\

	&	&No	&2.1001	&0.3862	&0.0117	&\\
\multirow{2}{*}{0.1}	&\multirow{2}{*}{50}	&Yes	&2.9221	&0.1701	&0.0128	&\multirow{2}{*}{139.13$\%$}\\

	&	&No	&2.9537	&0.1719	&0.0092	&\\

\end{tabular}
\end{ruledtabular}
\end{table*}

The width of the fixed-point number in this paper is smaller, so less storage resources (under 50$\%$) on-chip (BRAM) are consumed in our scheme even if the parallelism parameter $p$ is higher. Hence, we can instantiate two decoders on the FPGA. To improve the decoding performance, we add a residue error-bits erase module to decrease the FER. 

\subsection{\label{sub:level3}Computation complexity and security}
For the residue error-bits erase module, the resource consumption is mainly used for computing syndromes and the row weight of some matrices and all of them can be implemented by doing shift operation and XOR. Therefore, the overall computational complexity of this module is low. For the rate-0.2 LDPC code, the consumption of clock cycles of the residue error-bits erase module is approximately equal to 2.4 decoding iterations. For the rate-0.1 LDPC code, the consumption is approximately equal to 3 decoding iterations. Compared with the BP decoding, the residue error-bits erase module has smaller influence on the throughput of the proposed decoder. Consequently, the residue error-bits erase module is effective and efficient to reduce the FER. 

Besides, the residue error-bits erase module just works on the decoding results and there is no extra information exchange. Hence, this module brings no extra security problem to the CV-QKD system.

\section{\label{sec:level5}INFLUENCE FOR SECRET KEY RATE}
Secret key rate $K_t$  is the key indicator for QKD systems. The secret key rate per pulse of a CV-QKD system is given by \cite{r33,r34,r35} 
\begin{equation}\label{eq9}
\begin{aligned}
K_t= (1-FER)(\beta{I_{AB}}-\chi{_{BE}}),
\end{aligned}
\end{equation}
where $\beta$ is the reconciliation efficiency, $I_{AB}$ is the mutual information between two participants Alice and Bob, and $\chi_{BE}$ is the mutual information between Bob and Eve. As explained in \cite{r36}, $I_{AB}$ and $\chi_{BE}$ can be described as the function of the modulation variance $V_{A}$, that is, ${I_{AB}} =f_{I_{AB}}(V_{A}) $ and  $\chi_{BE} =f_{\chi_{BE}}(V_{A}) $. Besides, given the error correction codes and  the algorithm of information reconciliation, the reconciliation efficiency can be described as the function of $V_{A}$. 
\begin{equation}\label{eq10}
\begin{aligned}
\beta=\frac{R}{0.5\log_{2}{(1+SNR)} }=f_{\beta}(V_A). 
\end{aligned}
\end{equation}

The blue curve in Fig.~\ref{fig3} shows the relationship between FER and SNR for the rate-0.2 QC-LDPC code. Because SNR is the function of $V_{A}$, we can get the fitting function of $f_{FER}(V_A)$ as given in \cite{r36}. Thus, the secret key rate $K_t$ can be written  as  
\begin{equation}\label{eq11}
\begin{aligned}
K_t=(1-f_{FER}(V_A))(f_{\beta}(V_A)f_{I_{AB}}(V_A)-f_{\chi_{BE} }(V_A)).
\end{aligned}
\end{equation}

From Eq.(~\ref{eq11}), we can get the maximum secret key rate $K_{opt}$ by optimizing the parameter $V_{A}$ . In a similar way, according to Fig.~\ref{fig6}, we get the fitting function $f_{FER}(V_A)$ for the rate-0.1 QC-LDPC code and its optimal secret key rate.

Table~\ref{tab:tab3} shows the optimization results of secret key rate by the method in \cite{r36} for the decoder with/without residue error-bits erase module. We observe that, with residue error-bits erase module, the optimal secret key rate is improved up to 425.64$\%$ and 139.13$\%$ respectively, compared with that without error-bits erase module. In fact, the real-time throughput of CV-QKD is determined by the error correction throughput, the repetition rate can be increased to 360.92 pulse/s and 194.65 pulse/s correspondly. As a consequence, the proposed method in this paper can support 17.97 Mbps and 2.48 Mbps real-time generation of secret key rate under transmission distance of 25km and 50km (multiplying the optimal secret key rate and the repetiton rate).  
\section{\label{sec:level6}CONCLUSION}
In this paper, a FPGA-based LDPC decoder with limited precision for CV-QKD is proposed. Fixed-point number with fewer bits is adopted to reduce the consumption of hardware resources. A residue error-bits erase module is used to reduce the FER caused by finite precision of the decoder. For the codes of rates 0.2 and 0.1, 425.64$\%$ and 139.13$\%$ improvement of secret key rates are achieved compared with that without the residue error-bits erase module, and the corresponding decoder throughputs are 194.65Mbps and 360.92 Mbps, which can support 17.97 Mbps and 2.48 Mbps real-time generation of secret key rate for CV-QKD systems under distances of 25km and 50km. The method proposed in this paper provides an effective and efficient method to support high-speed real-time CV-QKD, and paves the way for high-rate real-time CV-QKD deployment in secure metropolitan area network.

\begin{acknowledgments}
This work was supported in part by the National Natural Science Foundation of China (Grant Nos 61901425, U19A2076, 62171418 and 62101516), the National Key Research and Development Program of China (Grant No. 2020YFA030970X), the Technology Innovation and Development Foundation of China Cyber Security (Grant No. JSCX2021JC001), the Chengdu Major Science and Technology Innovation Program (Grant No. 2021-YF08-00040-GX), the Chengdu Key Research and Development Support Program (Grants Nos 2021-YF05-02430-GX and 2021-YF09-00116-GX), the Sichuan Science and Technology Program (Grants Nos 2022YFG0330 and 2021YJ0313) and the Foundation and Advanced
Research Projects of Chongqing Municipal Science and Technology Commission
(China) under Grant cstc2019jcyjmsxmX0110.
\end{acknowledgments}

\appendix

\section{\label{app1}Memory arrangement}
\begin{figure}[htb]
\centerline{
\includegraphics[width = 5cm]{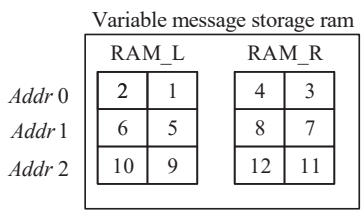}}% Here is how to import EPS art
\caption{\label{fig7} Structure of the variable message storage RAM, the number 1,2,3,...,12 correspond to the column index of \emph{\textbf{H}}.}
\end{figure}

The variable message storage RAM consists of two RAMs: RAM$\_$L and RAM$\_$R. We use $N$ to represent the code length and $p$ to represent the parallelism of the proposed architecture (the result of $N/p$ is an even number). According to the structure of \emph{\textbf{H}} in Fig.~\ref{fig1}, Fig.~\ref{fig7} shows the structure of the variable message storage RAM with $N$ = 12 and $p$ = 2.

As shown in Fig.~\ref{fig7}, the number of $L_{q_{n}}$ that is stored in an address of both RAM$\_$L and RAM$\_$R is $p$, so 2$p$ data can be read from the variable message storage RAM in a clock cycle. Then, a cyclic shift module presented in Appendix~\ref{app3} outputs $p$ data that is needed in next clock cycle from these 2$p$ data. The check message storage RAM consists of a RAM, and the structure of this RAM is as the same as RAM$\_$L, but with different depth. The bit width of Syn$\_$Ram is $p$ and the depth of Syn$\_$Ram is as the same as the check message storage RAM. Because \emph{\textbf{s}} is required in both Check node unit and Decode decision unit, to save the resource of Block Ram, the Syn$\_$Ram is configured as a true dual port RAM.
\section{\label{app2}Addr$\_$gen unit}
After finishing storing the receiving message into RAM$\_$L and RAM$\_$R, the iterative decoding starts. GCU uses signal $start$ and $stop$ to control the decoding process. When the value of $Iter$\_$num$ is equal to the maximum number of iterations, the iterative decoding  stops. According to Fig.~\ref{fig2}, $L_{q_{n}}$ and $L_{r_{mn}}$ are the inputs of the VNU. Because pipeline structure is adopted in designing this decoder, $L_{q_{n}}$ and $L_{r_{mn}}$ need to be read from the variable message storage RAM and the check message storage RAM constantly. $L_{r_{mn}}$ is read one by one, so the read address of check message storage RAM is generated by a simple counter. The read address of $L_{q_{n}}$ is computed according to the parity check matrix \emph{\textbf{H}} and is stored in a ROM, so the read address of variable message storage RAM is read from this ROM one by one.
\begin{table*}[htb]
\caption{\label{tab:tab4}Optimization of rate-0.1 and rate-0.2 MET-LDPC codes over the BI-AWGN channel. $\emph{\textbf{L(r,x)}}$ is the variable node degree distribution and $\emph{\textbf{R(x)}}$ is the check node degree distribution. $\sigma$ is the code threshold that represents the maximum channel noise level at which the decoding error probability converges to zero as the code length goes to infinity.}
\begin{ruledtabular}
\begin{tabular}{ccc}
$R$  &Degree distribution &$\sigma$\\ \hline
\multirow{2}{*}{0.1}	
&$$\emph{\textbf{L(r,x)}}$ =0.075r_1x^2_1x_2^{21}+0.05r_1x_1^3x_2^{20}+0.875r_1x_3^{1}$&\multirow{2}{*}{2.56}\\
&	$$\emph{\textbf{R(x)}}$=0.025x_1^{12}+0.825x_2^3x_3^1+0.05x_2^2x_3^1$	&\\
\multirow{2}{*}{0.2}	&$$\emph{\textbf{L(r,x)}}$ =0.1244r_1x^2_1x_2^{8}+0.1253r_1x_1^3x_2^{10}+0.7503r_1x_3^1$&\multirow{2}{*}{1.71}\\
&	$$\emph{\textbf{R(x)}}$=0.0341x_1^{12}+0.0156x_1^{11}+0.027x_2^2x_3+0.7476x_2^3x_3$	&\\
\end{tabular}
\end{ruledtabular}
\end{table*}
\section{\label{app3}Shift-right unit}
\begin{figure}[htb]
\centerline{
\includegraphics[width = 8cm]{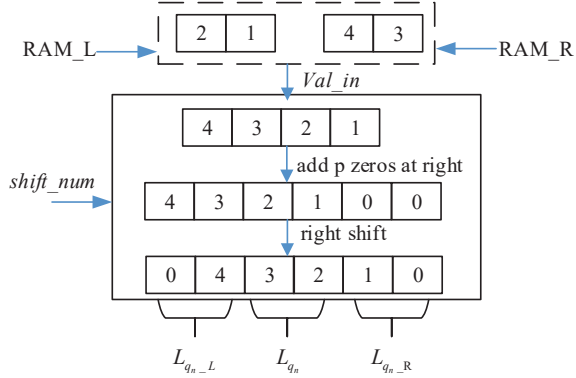}}% Here is how to import EPS art
\caption{\label{fig8}The structure of Shift-right unit.}
\end{figure}

Parameter $p$ indicates that this decoder needs to deal with $p$ data in a clock cycle. However, 2$p$ data are read from the variable message storage RAM. In this decoder, the Shift-right unit is designed to get $p$ data from these 2$p$ data. For example, as shown in Figs.~\ref{fig1} and \ref{fig7}, $p$ = 2 and $L_{q_{n}}$ with index 2 and index 3 (we use (2, 3) to represent) are needed at the beginning of the iterative decoding. Data (1, 2) and (3, 4) are read from RAM$\_$L and RAM$\_$R. The process of this unit to get (2, 3) is demonstrated in Fig~\ref{fig8}.

In Fig. 8, $val$\_$in$ and $shift$\_$num$ are the inputs of this unit. There are three outputs, therein, $L_{q_{n}}$ is used as the input of VNU, and $L_{q_{n}\_L}$ and $L_{q_{n}\_R}$ are stored in the FIFO buffer$\_$1. The reason why adding $p$ zeros at right is to make sure the three outputs have the same bit width. 

\section{\label{app4}Shift-left unit}
\begin{figure}[htb]
\centerline{
\includegraphics[width = 8cm]{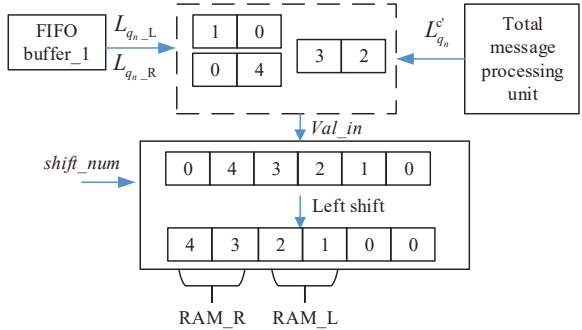}}% Here is how to import EPS art
\caption{\label{fig9}The structure of Shift-left unit.}
\end{figure}
The total message processing unit executes Eq.(~\ref{eq4}) and gets the output $L_{q_{n}}^{'}$. Then, $L_{q_{n}}^{'}$ will be written back to the RAM$\_$L and RAM$\_$R in right order by using the Shift-left unit. The structure of this unit is similar to the Shift-right unit. As shown in Fig.~\ref{fig9}, there are three inputs and two outputs in this unit. The two outputs are written back to RAM$\_$L and RAM$\_$R. There is clock delay between reading data from variable message storage RAM and writting them back, which results in the data that has been updated is covered. For example, in Fig.~\ref{fig2} and Fig.~\ref{fig7}, when updating the nodes in the first layer, data (1, 2) and (3, 4) are read to update (2, 3). A few clock cycles later, data (1, 2) and (3, 4) are read again to update (4,1), At this moment, data (2, 3) has not been updated because of the clock delay. Hence, when data (1, 2) and (3, 4) are written back at the second time, data (2, 3) covers the result that has been written back at the first time.

We denote $k = z/p$, then the total message processing unit will generate $k$ values named $L_{q_{n}}^{1^{'} }$ , $L_{q_{n}}^{2^{'}}$  , …,  $L_{q_{n}}^{k^{'}}$  when updating the nodes in a layer. To avoid data coverage, the input of the shift-left unit is changed according to the value of $c$ as follows:

(1)$c=1$, $Val$\_$in$ = $\{$$L_{q_{n}\_L}$, $L_{q_{n}}^{1^{'} }$, $L_{q_{n}\_R}$  $\}$.

(2)$1<c<k$, $Val$\_$in$ = $\{$$L_{q_{n}\_L}$, $L_{q_{n}}^{c^{'} }$, $L_{q_{n}\_R}^{c-1^{'}}$  $\}$.

(3)$c=k$, $Val$\_$in$ = $\{$ $L_{q_{n}}^{1^{'} }$ , $L_{q_{n}}^{c^{'} }$,  $L_{q_{n}\_R}^{c-1^{'}}$   $\}$.

\section{\label{app5}Construction of QC MET-LDPC codes}
Different LDPC code ensembles have different decoding performance which can be characterized by the degree distribution. At first, we use the density evolution method to get the optimized degree distribution for MET-LDPC codes. Then, we use progressive edge growth algorithm to construct QC MET-LDPC codes according to the optimized degree distributions. The optimized results of rate-0.1 and rate-0.2 MET-LDPC codes are shown in Table~\ref{tab:tab4}.

%~(\ref{appa}), (\ref{appb}), and (\ref{appc}).

\nocite{*}

\bibliography{apssamp}% Produces the bibliography via BibTeX.

\end{document}